\documentclass[10pt, conference]{IEEEtran}
\IEEEoverridecommandlockouts
\usepackage{cite}
\usepackage{amsmath,amssymb,amsfonts}
\usepackage{algorithmic}

\usepackage{graphicx}
\usepackage{textcomp}
\usepackage{xcolor}
\def\BibTeX{{\rm B\kern-.05em{\sc i\kern-.025em b}\kern-.08em
    T\kern-.1667em\lower.7ex\hbox{E}\kern-.125emX}}
\usepackage{enumitem, hyperref}
\makeatletter
\def\namedlabel#1#2{\begingroup
    #2%
    \def\@currentlabel{#2}%
    \phantomsection\label{#1}\endgroup
}
\makeatother
\begin{document}

\title{Calculating Originality of LLM Assisted Source Code \\
}

\author{\IEEEauthorblockN{Shipra Sharma}
\IEEEauthorblockA{shiprashar@gmail.com}
\and
\IEEEauthorblockN{ Balwinder Sodhi}
\IEEEauthorblockA{\textit{Department of Computer Science and Engineering} \\
\textit{Indian Institute of Technology Ropar}\\
India \\
sodhi@iitrpr.ac.in}
}

\maketitle

\begin{abstract}
The ease of using a Large Language Model (LLM) to answer a  wide variety of queries and their high availability has resulted in LLMs getting integrated into various applications. LLM-based recommenders are now routinely used by students as well as professional software programmers for code generation and testing. Though LLM-based technology has proven useful, its unethical and unattributed use by students and professionals is a growing cause of concern. As such, there is a need for tools and technologies which may assist teachers and other evaluators in identifying whether any portion of a source code is LLM generated. 

In this paper, we propose a neural network-based tool that instructors can use to determine the original effort (and LLM's contribution) put by students in writing source codes. Our tool is motivated by minimum description length measures like Kolmogorov complexity. Our initial experiments with moderate sized (up to 500 lines of code) have shown promising results that we report in this paper.  
\end{abstract}


\section{Introduction}
With the advent of Large Language Models (LLM) models such as ChatGPT, several coding tasks have become easy to complete via use of such LLMs. Such tasks include programming assignments in courses, generating subroutines and code fragments for commonly encountered algorithmic tasks, and so on. For example, programming assignments in many Computer Science and Engineering (CSE) courses can be generated in large measure \cite{dwivedi3} via these models. It has become very difficult to detect by standard plagiarism detection tools such as Turnitin \cite{weisz5}, that such source code is LLM generated. Even a complex assignment can be broken into  simpler components, and each component can be written separately using such LLMs. Given this situation, it is highly desirable to construct a tool which can detect unauthorized or unattributed LLM help taken by the students in preparing their coding assignments. Usage of such LLM-assisted coding tools is recommended as the engineers/students may be required by the employers to be conversant with the use of such tools \cite{weisz5,peng6, khalil4, anu7}. 

Although the LLM-based coding assistant tools seem to reply correctly to complex queries akin to an expert, they still lack the conceptual understanding of the queries as well as the results generated by the tool. The major shortcoming of these tools is lack of deep reasoning and analytical skills \cite{rosenblatt2, khalil4, ss8}. Hence, before we begin to resolve the difficulties mentioned above, we should first be able to measure (at least approximately) the amount of originality in an assignment. Motivated by the above, and by potential applications in the domain of Software Engineering, we consider the following research questions in this paper.

\begin{description}
\item [\namedlabel{RQ1}{$RQ\: 1$}] \textit{ Can we quantify the amount of original contribution by a student in an assignment, assuming that he/she has used an LLM such as ChatGPT for its preparation?}

\item [\namedlabel{RQ2}{$RQ\: 2$}] \textit{ How can we detect the similarity in the original contribution portion of two separate submissions when it is known that the students can take assistance from LLM-based tools in creating the submissions?}


\item  [\namedlabel{RQ3}{$RQ\: 3$}] \textit{ How efficiently can we automate our answers to the above questions?}
\end{description}

In this paper, we propose two scores: the {\it originality score} $o(D)$ and the {\it similarity score} $s(D)$ of a source code $D$ as solutions to the above questions.
We further propose to use these scores extensively in an adaptable teaching process as follows:

\begin{enumerate}
\item Students with less measure of original contribution in their assignments (i.e., less originality scores) may be awarded suitably reduced scores. 
\item Students with large amounts of overlap in their respective contributions (i.e., high similarity scores) may not be awarded extra “originality credits”.
\item More credits may be allocated to the “difficult” fragments of the program (or, assignment submission), and lesser credits may be allocated to the “easier” fragments of the program (or, assignment submission).
\end{enumerate}

These steps will lead to a constructive assessment of students, which encourages the students to develop original and high-depth analytic thinking.

The above discussed scenario is one of the many applications of our work. Others are its usage in software development as these LLM-based models cannot replace software engineers (as of now), but can assist them \cite{weisz5, peng6}.

\section{Computing originality score of a program}

\subsection{Setting up the problem}
\label{problem-setup}
\begin{figure}
\includegraphics[scale=0.28]{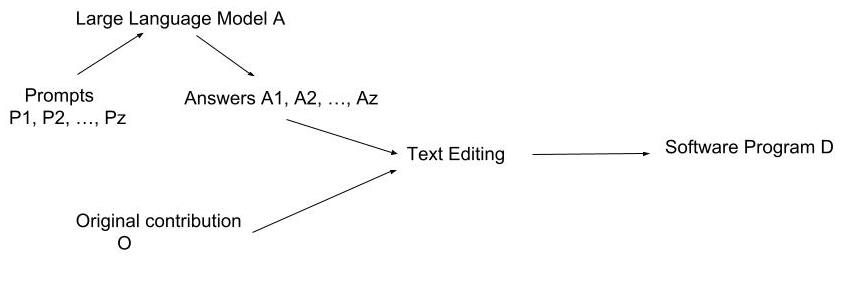}
\caption{Constructing a program $P$ using LLMs}
\label{fig1}
\end{figure}

Suppose a programmer has unlimited access to a large language model $\mathcal{A}$ ($\mathcal{A}$ can be ChatGPT, GPT-J, etc.). The programmer constructs a software program $D$ using (see Figure \ref{fig1}):

\begin{enumerate}
\item the answers $A_1, A_2, \ldots, A_z$ to a sequence $P_1, P_2, \ldots, P_z$ of $z$ {\it prompts} to $\mathcal{A}$, and
\item the programmer's own {\it original contribution} $\mathcal{O}$.
\end{enumerate}

Program $D$ is finally constructed by combining $A_1, A_2, \ldots, A_z$ and $\mathcal{O}$ using conventional text editing, rearrangements, etc. To be more specific, a conventional plagiarism detection software (say, Turnitin) will detect high similarity between the strings $D$ and the corpus $\{A_1, A_2, \ldots, A_n, \mathcal{O}\}$.

We define the following metrics:
\begin{itemize}
    \item {\it total effort} $e(D)$ of the programmer as the total length of all prompts and the programmer's original contribution: 
        $$e(D) = \sum_{i=1}^{z} |P_i| + |\mathcal{O}|$$ 

    \item {\it originality score} $o(D)$ ($0 \leq o(D) \leq 1$) of the program:
     $$o(D) = \frac{|\mathcal{O}|}{|D|}$$
    
\end{itemize}
 Our assumption is that a lower originality score would imply a lower original contribution by the programmer. Any programmer or student using LLM models to assist in writing programs implicitly minimizes $e(D)$ and in turn also minimizes $o(D)$. This motivates the following question.

{\it Question $1$. Given a document $D$ and LLM $\mathcal{A}$, calculate the minimum originality score $o(D)$.}
(This corresponds to \ref{RQ1}).

\subsection{Solving \ref{RQ1}}
\label{solve-problem}
To solve Question $1$ we bound the maximum number of prompts $z$, which is a positive integer and the maximum length $L$ of each prompt ($P_1, P_2, \ldots, P_z$). We now formulate a bounded version of Question $1$ above:\\

{\it Question $1.1$. Compute the minimum value of the originality score $o(D)$, under the assumption that the programmer can give at most $z$ prompts, each of length at most $L$.}\\

\begin{figure}[t]
\includegraphics[scale=0.24]{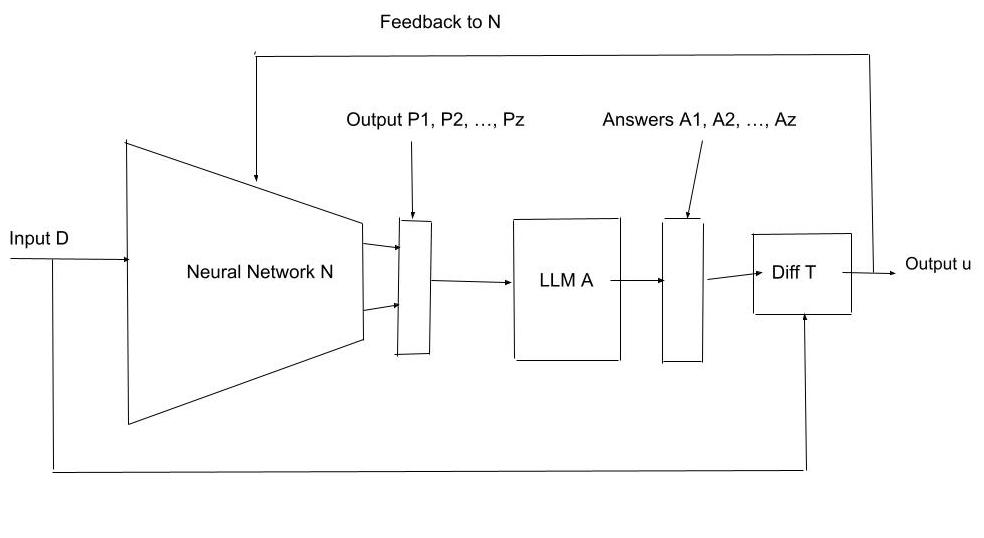}
\caption{Calculating originality score}
\label{fig2}
\end{figure}

Let $T$ be a conventional plagiarism detector (a trivial one to use could be the {\tt diff} command in UNIX-based systems). Figure \ref{fig2} illustrates the algorithm for solving Question $1.1$.

The program $D$ in Figure \ref{fig2} forms the input to a neural network $N$. The output of $N$ is of size $z \cdot L$, and corresponds to the $z$ unknown prompts to LLM $\mathcal{A}$. The output of $N$ is given as input to LLM $\mathcal{A}$ to obtain answers $A_1, A_2, \ldots, A_z$. A conventional plagiarism detector $T$ is used to find the similarity percentage $t$ between $D$ and the output answers $(A_1, A_2, \ldots, A_n)$. The original contribution $\mathcal{O}$ is estimated by removing the parts of $D$ which match with the output answers. Finally, the output (originality score) $u$ is equal to $\frac{|\mathcal{O}|}{|D|}$. If the similarity percentage between $D$ and $(A_1, A_2, \ldots, A_n)$ is $t$, the originality score is expected to be approximately $1 - 0.01 \cdot t$\footnote{as $t$ is percentage score we convert it to a number between $0$ and $1$ by multiplying by $0.01$}. The output originality score $u$ is given as the feedback to neural network $N$, with the objective of minimizing $u$. 

{\it Remark.} Please note that giving the same prompt again to an LLM can generate somewhat different answers. To cover all possibilities, our model allows for the same prompt to be repeated more than once in the sequence $P_1, P_2, \ldots, P_z$.

\subsection{Applying the minimum description length (MDL) principle}
\label{mdl-principle}

The {\it minimum description length (MDL) principle} \cite{mdl1} is a 
well-known principle for model selection. The MDL principle always selects the {\it shortest description} of given data, from the set of all possible descriptions. The quantity $\Gamma=(P_1, P_2, \ldots, P_z, \mathcal{O})$ (see Section \ref{problem-setup}) can be viewed as the content comprising of prompts plus the original code added by the student that results in the desired program as the output from an LLM. Thus, $\Gamma$ can be thought to represent a {\it description} of $D$, which can lead to generation of the desired code. In other words, given the description $\Gamma$ and LLM $\mathcal{A}$, we can reconstruct program $D$ almost completely.

Our proposed solution (see Section \ref{solve-problem}) can then be viewed as an application of the MDL principle. For each possible description $\Gamma$, our algorithm selects the description with minimum ``length", where the length of a description $\Gamma$ is defined as its originality score $\frac{|\mathcal{O}|}{|D|}$.

\section{Computing similarity score of two programs}

\subsection{Setting up the problem}

Suppose two programmers Alice and Bob produce programs $D_1$ and $D_2$ respectively. Both programs solve the same computational problem, and both Alice and Bob had unlimited access to LLM $\mathcal{A}$ during the coding process.

Suppose Alice constructed $D_1$ using prompts $P_1, P_2, \ldots, P_z$ and original contribution $\mathcal{O}_1$. Similarly, suppose Bob constructed $D_2$ using prompts $Q_1, Q_2, \ldots, Q_z$ and original contribution $\mathcal{O}_2$. Let $p$ be the similarity percentage between the two descriptions, $\Gamma_1=(P_1, P_2, \ldots, P_z, \mathcal{O}_1)$ and $\Gamma_2=(Q_1, Q_2, \ldots, Q_z, \mathcal{O}_2)$ using the conventional plagiarism detector $T$.

Then we define similarity score, $$s(D_1, D_2) = 0.01 \cdot p$$ 

We now state the second question considered in this paper:\\

{\it Question $2$. Given two source codes $D_1$ and $D_2$ and LLM $\mathcal{A}$, calculate the similarity score $s(D_1, D_2)$.} (This corresponds to \ref{RQ2}.)

\subsection{Solving \ref{RQ2}}

In analogy with our approach for originality score, we consider a bounded version of Question $2$:\\

{\it Question $2.1$. Given two source codes $D_1$ and $D_2$, compute the maximum value of similarity score $s(D_1, D_2)$, under the assumption that both Alice and Bob can give at most $z$ prompts, each of length at most $L$.}\\

Figure \ref{fig3} illustrates the algorithm for solving Question $2.1$:

\begin{figure}
\begin{center}
\includegraphics[scale=0.245]{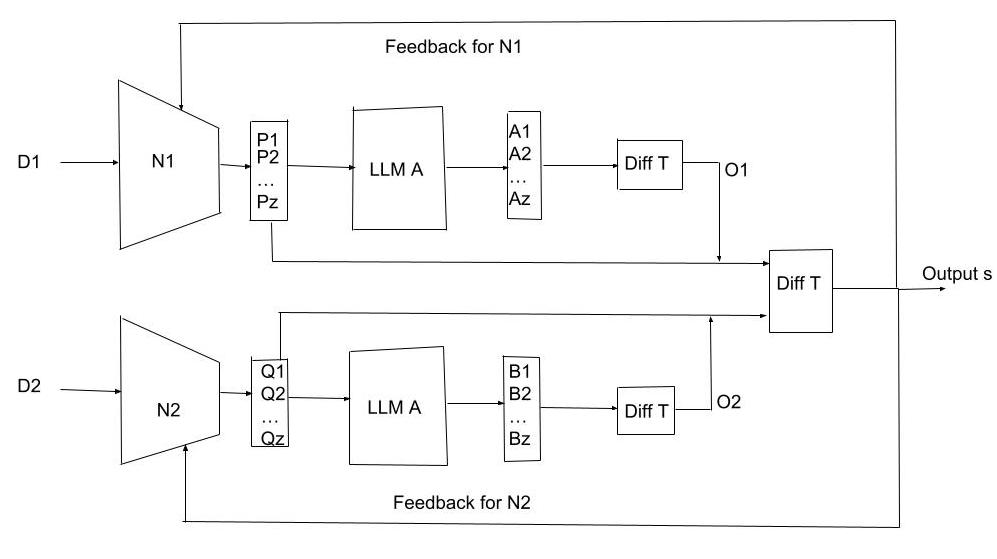}
\caption{Calculating similarity score}
\label{fig3}
\end{center}
\end{figure}

Source codes $D_1$ and $D_2$ are the inputs to two neural networks $N_1$ and $N_2$. The output of each neural network is of size $z \cdot L$. The output of $N_1$ corresponds to the $z$ unknown prompts of Alice and the output of $N_2$ corresponds to the $z$ unknown prompts of Bob. Next, the outputs of $N_1$ and $N_2$ are given as input to LLM $\mathcal{A}$ to generate answers $A_1, A_2, \ldots, A_z$ and $B_1, B_2, \ldots, B_z$ respectively.

Using algorithm $T$, we compute the original contribution $\mathcal{O}_1$ of Alice for prompts $P_1, P_2, \ldots, P_z$ and the original contribution $\mathcal{O}_2$ of Bob for prompts $Q_1, Q_2, \ldots, Q_z$. Finally, the similarity $s$ between $(P_1, P_2, \ldots, P_z, \mathcal{O}_1)$ and 
$(Q_1, Q_2, \ldots, Q_z, \mathcal{O}_2)$ is computed using $T$, and this is used as feedback for both neural networks $N_1$ and $N_2$. The objective of the training process is to maximize (see Question $2.1$) the output similarity $s$. 

{\it Remark $1$.} In our implementation, we input $(D_1, D_2)$ to a single neural network $N$, with ouput $(P_1, P_2, \ldots, P_z, Q_1, Q_2, \ldots, Q_z$). The intuition is that a single neural network may lead to faster convergence due to information flow along cross connections between input neurons of $D_1$ and $D_2$.

{\it Remark $2$.} In terms of MDL principle, the above network tries to compute the shortest description $((P_1, P_2, \ldots, P_z, \mathcal{O}_1), (Q_1, Q_2, \ldots, Q_z, \mathcal{O}_2))$ of $(D_1, D_2)$, where the ``length" of the description is defined as the similarity score of $T$ on inputs $(P_1, P_2, \ldots, P_z, \mathcal{O}_1)$ and $(Q_1, Q_2, \ldots, Q_z, \mathcal{O}_2)$.

\section{Previous Work}

{\it Kolmogorov complexity and related measures.} When the algorithm $\mathcal{A}$ is a universal Turing machine (instead of a LLM), the minimum length description of program $P$ is called its Kolmogorov complexity \cite{kolmogorovbook}. In \cite{goldblum2023free}, the authors propose that neural network models such as GPT-3 have a ``simplicity bias" and prefer data with low Kolmogorov complexity. Kolmogorov complexity inspired measures have a long history of application in similarity detection and compression. In \cite{livitanyi1}, the authors define a similarity metric called Normalized Information Distance (NID), based on Kolmogorov complexity. Since Kolmogorov complexity is non-computable, the authors further develop the notion of Normalized Compression Distance (NCD), which is an efficiently computable variant of NID using compression algorithms like gzip. More in-depth treatment of this topic is available in \cite{vitanyi2, vitanyi3, cilibrasi2} and related papers.\\

{\it Autoencoders.} An autoencoder \cite{autoencoder} is a neural network which first compresses the input using an encoder network and then tries to recover the input from the compressed code by using a decoder network \cite{deeplearningbook}. For the use of minimum description length (MDL) principle for autoencoders, see \cite{hinton}. In the algorithm proposed in this paper (Figure \ref{fig2}), the neural network $N$ can be viewed as the encoder, and the LLM $\mathcal{A}$ can be viewed as the decoder. Further, note that only the encoder is trained using feedback from the output.\\

{\it AI-detection tools.} We briefly discuss few recent softwares for detecting whether a
text is generated by a LLM or written by a human. An AI text classifier by OpenAI, the company behind ChatGPT, is now available \cite{openai-classifier}. The classifier outputs the probability that a given input text is AI-generated. GPTZero \cite{gptzero} is another AI-detection tool, which also provides scores for burstiness and perplexity \cite{burstiness, perplexity}. Another well-known tool is Originality.AI \cite{originalityai}.

\section{Preliminary experiments and vision for future work}

For an initial experimental setup for the proposed ideas, we designed a {\it prompt space} $\mathcal{P}$ of size $64$. Each prompt in this space is defined by a tuple of three words taken from independent sets $A, B, C$. Each of $A, B$ and $C$  contains words taken from common programming vocabulary encountered while describing the programs. For our experiments we chose $|A|=8, |B|=2, |C|=4$. For example, if the prompt is (``insertion", ``sort", ``C"), it is equivalent to writing a prompt: \texttt{Write the code for insertion sort in C}. We generated a pool of $10$ answers to this prompt using calls to ChatGPT and BLOOM. BLOOM model was run on Macintosh, while ChatGPT was prompted through API calls. This gave us a collection of $64 \cdot 10 = 640$ $(prompt, answer)$ pairs. We store this set in an offline repository $\mathcal{R}$ which we used to train a neural network $N$ using PyTorch. For each $answer$ the neural network was trained with the following loss function: {\it generate two prompts independently at random from the output probability distribution and calculate their similarity with $answer$}.

Next, we collected a test set $\mathcal{T}$  of $50$ programs. Each program $D$ in $\mathcal{T}$ was manually evaluated for similarity with the repository. Accordingly, an originality score $o(D)$ was assigned to every program in $\mathcal{T}$ using the formulas discussed in Section \ref{problem-setup}.

The neural network $N$ takes as input a source code $D\in \mathcal{T}$ and the output is a probability distribution over the prompt space $\mathcal{P}$. The best score provided by the neural network is the computed originality score $f(D)$ for two prompts. We found that the mean squared error $\epsilon$ between $o(D)$ and $f(D)$ was $0.3$ ($0\leq \epsilon \leq 1$), which is an encouraging result (\ref{RQ3}) .

This experiment required a considerable amount of manual effort as our goal was to prove the viability of our proposed idea. As the proposed idea shows to be implementable and valid, we propose the following research vision:

\begin{enumerate}

\item We plan to create a prompt space that accurately maps with the internal representation of prompts for large-scale deployed LLMs such as BLOOM, ChatGPT, BARD etc.
\item We plan to increase the size of repository $\mathcal{R}$, so that it consists of a realistic number of  $(prompt, answer)$ pairs. 
\item In future we plan to automate data cleaning, processing and model building so that the model can be trained and updated on real world data on regular basis.
\item We plan to increase the number of prompts in the prompt sequence to at least $20$.
\item Finally, we will define prompt complexity, and how it minimizes originality score to be always less then 0.45. The implication being that easier the prompt is to write to get the desired code fragment., lesser will be the originality score of a source code.


\end{enumerate}

%



\section{Conclusion}
As current plagiarism detection tools use a corpus of documents obtained from various sources for comparison, we {\it envision an originality detection tool} which generates a prompt sequence and calculates the minimum originality score. The key idea we have proposed in this paper is: the tools for detecting originality of LLM generated source code need to ``learn'' from the LLM generated source code itself and the prompts used to generate such source code.

Rather than trying to compute the probability that a text is AI-generated or human-generated (this has its technical limitations), we feel the focus should be on computing originality score using a pool of LLMs.

Our initial results are encouraging, and our computed originality scores are in agreement with human evaluations of originality and similarity.

\end{document}